\documentclass[12pt]{iopart}
\usepackage[utf8x]{inputenc}
\usepackage{graphicx, array}
\usepackage[caption=false]{subfig}
\usepackage[T1]{fontenc}

\begin{document}
\title[Relaxation of moir\'e patterns for slightly misaligned identical lattices]{Relaxation of moir\'e patterns for slightly misaligned identical lattices: graphene on graphite}

\author{M M van Wijk, A Schuring, M I Katsnelson and A Fasolino}

\address{Radboud University Nijmegen, Institute for Molecules and Materials, Heyendaalseweg 135, 6525 AJ Nijmegen, The Netherlands}
\ead{a.fasolino@science.ru.nl}

\begin{abstract}
We study the effect of atomic relaxation on the structure of moir\'e patterns in twisted graphene on graphite and double layer graphene by large scale atomistic simulations. The reconstructed structure can be described as a superlattice of `hot spots' with vortex-like behaviour of in-plane atomic displacements and increasing out-of-plane displacements with decreasing angle. These lattice distortions affect both scalar and vector potential and the resulting electronic properties. At low misorientation angles (<$\sim$1$^\circ$) the optimized structures deviate drastically from the sinusoidal modulation which is often assumed in calculations of the electronic properties. The proposed structure might be verified by scanning probe microscopy measurements.
\end{abstract}

\maketitle

\section{Introduction}
The study of van der Waals heterostructures, artificial materials made of stacked layers of different two dimensional (2D) materials, is a very promising new direction in nanoscience and nanotechnology~\cite{geimgrigorieva}. Apart from perspectives in applications, such as vertical geometry tunneling transistors~\cite{britnell2012field}, these systems are important for fundamental science. Graphene on hexagonal BN (h-BN), in particular, turns out to be an ideal playground to study statistical and quantum mechanics in tunable incommensurate potentials~\cite{yankowitz2012emergence, ponomarenko2013cloning, dean2013hofstadter, hunt2013massive,woods2014commensurate,ourPRL}. When two lattices with different lattice constants or different orientations are stacked upon each other, a larger periodicity, known as a moir\'e pattern, emerges. For graphene on h-BN, the moir\'e pattern depends on both lattice constant mismatch and misorientation. It was recently shown both experimentally and theoretically that in this case an incommensurate-to-commensurate transition for small misorientations takes place~\cite{woods2014commensurate,ourPRL}.

The case of identical lattices where misorientation is the only source of incommensurability may be physically different.  The case of large misorientation ($\sim 30^\circ$) was studied intensively for graphite flakes on graphite after recognizing the drastic reduction in sliding friction often called superlubricity~\cite{dienwiebel2004superlubricity}. Conversely, stick-slip dynamics with high friction were observed for small misorientation angles ($<5^\circ$) and these angles were therefore less studied in this context. The case of small angles became interesting after experimental studies of misaligned double layer graphene~\cite{li2010observation,luican2011single} because of the strong effect on the electronic structure, such as the appearance of secundary Dirac cones and van Hove singularities~\cite{bistritzer2011moire, trambly2010localization,schmidt2014superlattice,chen2014raman}. In most theoretical works~\cite{bistritzer2011moire,wallbank2014moire} the moir\'e patterns were modeled by rigid misaligned layers and the effective potential on the electrons was considered as the superposition of crystalline potentials from the two rotated layers. 

In one dimension (1D), starting from the seminal papers by van der Merwe~\cite{vandermerwe}, many studies~\cite{bak1982} have shown that the superposition of two slightly incommensurate periodicities lead to the appearance of a soliton lattice where commensurate (phase-locked) regions are separated by thin misfit regions. The 2D case  is much less studied but a qualitatively similar behaviour,  where vortices in 2D play the role solitons in 1D, was found in models of adsorbed monolayers on surfaces~\cite{snyman1981computed}. For the case of twisted double layer graphene, a density functional theory (DFT) calculation down to angles $<1^\circ$ has been recently reported~\cite{uchida2014atomic}. For the case of very small angles, a full DFT based minimization of the lattice was not performed in view of the diverging size of the supercell. It was assumed that the extrapolation to small angles of the calculated sinusoidal modulation for angles~$>3^\circ$, could describe the out-of-plane distortions. 

Classical simulations for carbon have been shown to describe structural properties accurately~\cite{brenner2002second,los2005improved}. They allow to study the very large samples needed for very small misalignment and can be used to verify the nature of the modulation in this limit. Here we use an atomistic model based on the REBO~\cite{brenner2002second} and Kolmogorov-Crespi~\cite{KC2005} potentials to optimize the structure of graphene on graphite and of double layer graphene as a function of the misalignment. 

We find an essential reconstruction of the geometrical moir\'e pattern at very small angles. The modulation cannot be described by a sinusoidal function anymore as often assumed~\cite{uchida2014atomic}. Instead, a 2D lattice of misfit regions (`hot spots') with large out-of-plane displacements separated by flat domains is formed. 

\section{Model} 
\begin{figure}
        \centering
        \subfloat[]{
          \includegraphics[width=0.3\linewidth]{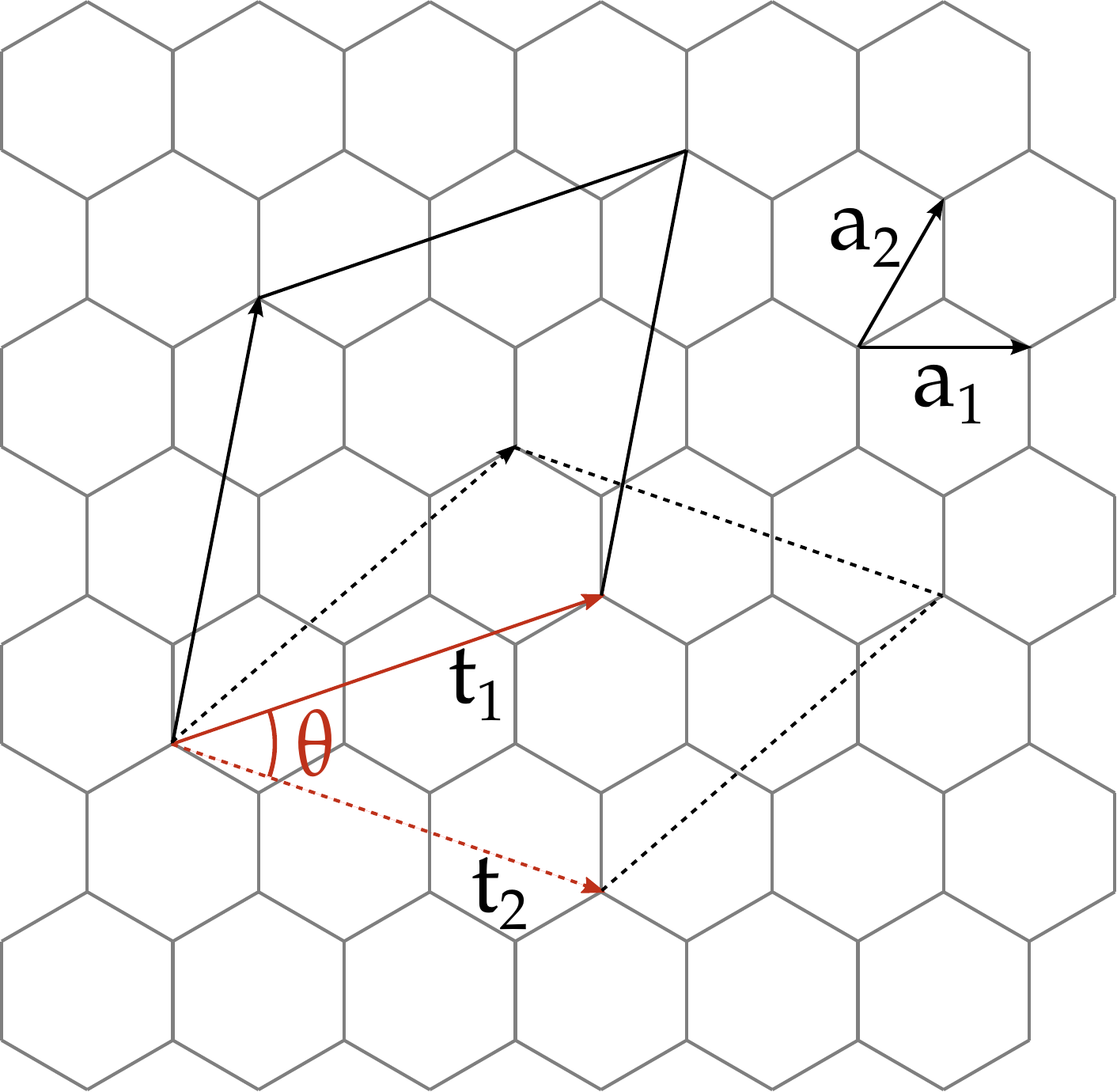}
        }
	\hspace{0.05\linewidth}
	\subfloat[]{
           \includegraphics[width=0.45\linewidth]{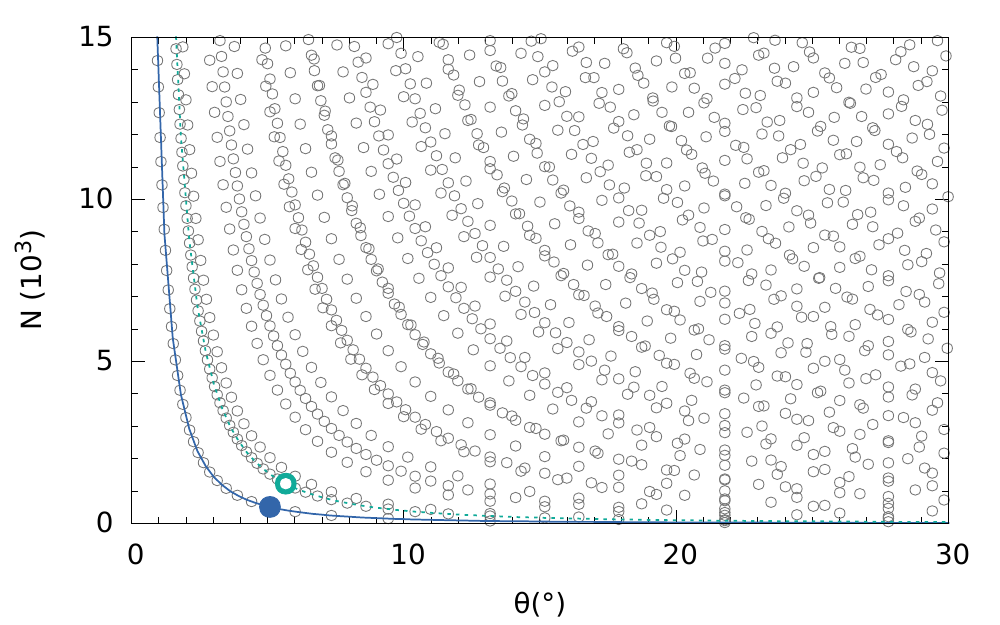}
        }\\
        \subfloat[]{
          \includegraphics[width=0.5145\linewidth]{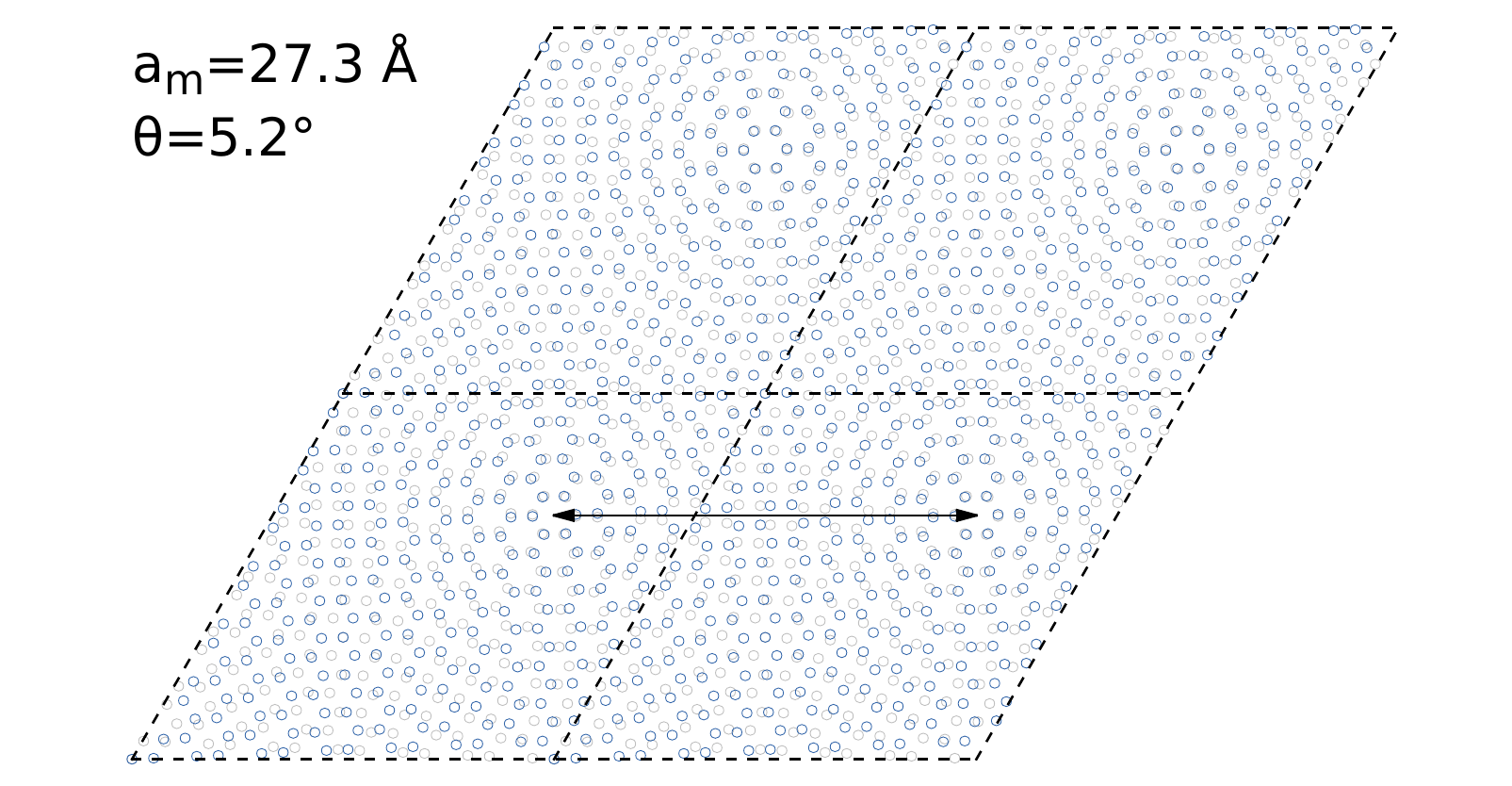}
	}
	\hspace{-0.05\linewidth}
        \subfloat[]{
           \includegraphics[width=0.4\linewidth]{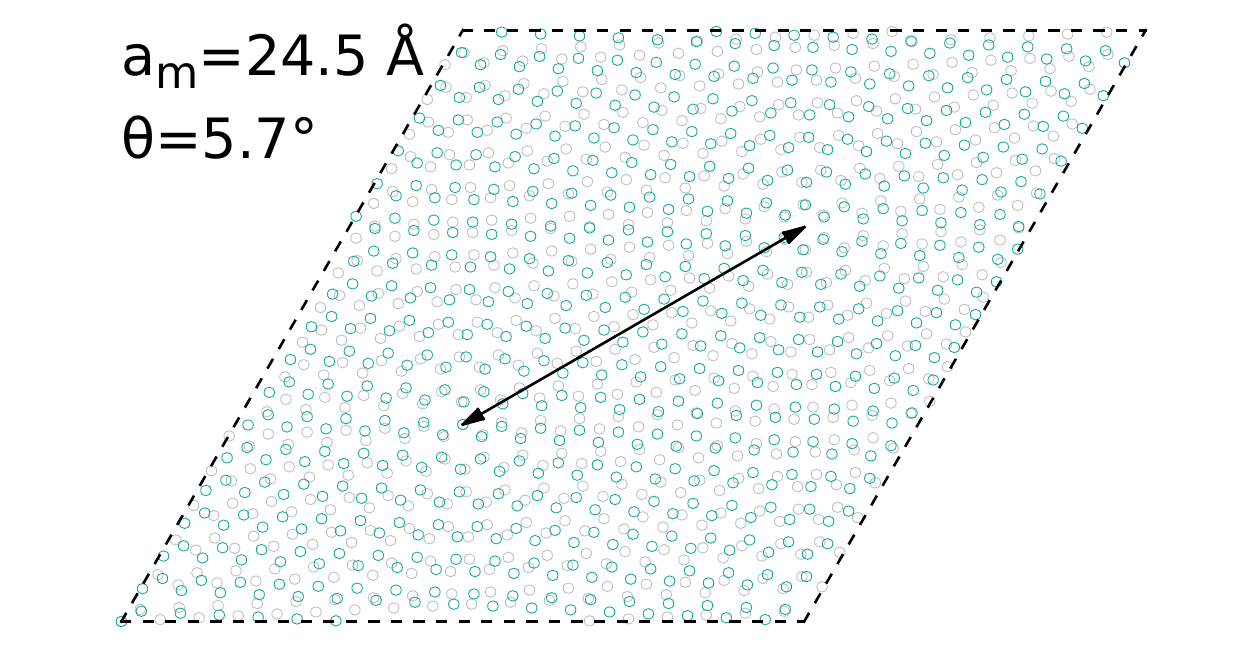}
        }
        \caption{(a) Schematic  construction of the unit cell. (b) Number of atoms $N$ for $N<$15000, $n<500$ and $m<n$, mirrored around 30$^\circ$. The dark blue line shows $N_m$ and the dashed light blue line 3$N_m$. (c) Four supercells with one moir\'e pattern in each of them ((n,m)=(7,6), $N_m$ indicated by the filled dot in (b)) (d) A single supercell with three moir\'e patterns in it ((n,m)=(17,1), $N_m$ indicated by the empty dot in (b)). Note that a similar value of the distance between moir\'e patterns $a_m$, indicated by the arrow, can correspond to different supercells. }
	\label{fig:schemeandN}
\end{figure}

When two lattices with different lattice constants or different orientations are stacked upon each other, a larger periodicity emerges: a moir\'e pattern. The lattice mismatch and relative orientation determine the size of the moir\'e pattern. For the case of graphene on graphite, the lattice constants are equal and the size of the moir\'e pattern can be written as~\cite{hermann2012periodic,shallcross2010electronic}
\begin{equation}
 a_{m}=\frac{a_{lattice}}{2 | \sin (\theta/2) |}
\end{equation}
where $\theta$ is the relative orientation of two layers, $a_{lattice}=\sqrt{3}d$ with $d$ the distance between carbon atoms. In \fref{fig:schemeandN}c,d $a_{m}$ is indicated by an arrow between the centers of the moir\'e patterns.

One may construct samples of twisted graphene satisfying periodic boundary conditions in the in-plane $x,y$ directions by identifying a common periodicity~\cite{shallcross2010electronic,savini2011bending,KC2005}  in the two rotated layers as illustrated in \fref{fig:schemeandN}a. For one layer, we define  a supercell with a basis vector $t_1=n a_1 + m a_2$ where $a_1$ and $a_2$ are the basis vectors of the graphene unit cell shown in \fref{fig:schemeandN}a with $n$ and $m$ integers and $n>m\ge 1$. For the second layer, a cell with same size and rotated by an angle $\theta$ can be obtained  by taking a basis vector $t_2=(n+m) a_1 -m a_2$. The common supercell is then obtained by rotating by $\theta/2$ the cell with basis vector $t_1$ and  by $-\theta/2$  the cell with basis vector $t_2$. Each pair of n and m thus identifies a common supercell for two layers rotated by an angle $\theta$ given by 
\begin{equation}
 \cos\theta=\frac{2n^2+2nm-m^2}{2(n^2+nm+m^2)}.
\end{equation}
 The supercell has sides of length 
\begin{equation}
 a_{cell}=|t_1|=|t_2|=a_{lattice}\sqrt{n^2+nm+m^2}
\end{equation}
 and contains  $N$  atoms, with
\begin{equation}
 N=4(n^2+nm+m^2).
\end{equation}
For small angles, this procedure quickly results in very large supercells, see \fref{fig:schemeandN}b.

The size and shape of the moir\'e pattern does not change if the top layer shifts with respect to the bottom layer. In fact, a 60$^\circ$ rotation corresponds to a change from AA to AB stacking which can also be achieved by a translation. The symmetry of the hexagonal lattice combined with the translational freedom lead to a symmetry around $30^\circ$.

The number of atoms in a single moir\'e pattern is given by~\cite{hermann2012periodic}
\begin{equation}
N_m=\frac{1}{\sin^2\theta}.
\end{equation}
This value also forms a lower bound for the number of atoms in the supercell since each supercell contains an integer number of moir\'e patterns. The values of $N$  on the minimum bound line in \fref{fig:schemeandN}b correspond to supercells with $n-m=1$, like the one in \fref{fig:schemeandN}c. The next highlighted  line in \fref{fig:schemeandN}b corresponds to 3$N_m$, for supercells with $m=1$ and three moir\'e patterns inside them (see \fref{fig:schemeandN}d). The three  moir\'e patterns in the supercell are not exactly equal due to the discreteness of the lattice, in the same way as  a shift of the top layer leads to a moir\'e pattern that is not exactly the same as prior to the shift. In \fref{fig:schemeandN}d one can see that the center of one moir\'e pattern is on an atom while the other is on the center of a hexagon. However, the differences between the moir\'e patterns are tiny and cannot be distinguished experimentally~\cite{yildiz2015apparent}.

The size of the moir\'e pattern is thus determined by geometry, but its structure  is determined by the atomic displacements that minimize the total energy.  We have studied the effects of relaxation on the structure of  the moir\'e patterns of twisted graphene on a graphite substrate and of double layer graphene. The interactions between atoms within the layer are given by the \textsc{REBO} potential~\cite{brenner2002second} as implemented in the molecular dynamics code \textsc{LAMMPS}~\cite{lammps}. The interlayer interactions are given by the registry-dependent Kolmogorov-Crespi potential without the normals~\cite{KC2005}, scaled to match the equilibrium distance $d=$1.3978~\AA~of the intralayer potential. The combination of these potentials has been shown to accurately reproduce the potential energy surface of the graphite substrate~\cite{reguzzoni2012potential}, although the corrugation is possibly underestimated~\cite{lebedeva2011interlayer}.

Using the damped dynamics algorithm \textsc{FIRE}~\cite{bitzek2006fire} we relax  the atomic positions to their minimum energy configuration. As a model of graphene on graphite we consider a mobile layer of graphene on a rigid substrate whereas to model double layer graphene we allow atomic relaxations in both layers. 
For graphene on graphite we consider not only full relaxation in all directions but also a case where the interlayer distance $z$ is kept fixed at 3~\AA. As this value is smaller than the equilibrium value (3.32~\AA~for AB stacking), the potential energy corrugation between layers is enhanced.

\section{Graphene on graphite}
\begin{figure}
        \centering
        \subfloat[]{
                \includegraphics[width=0.37\linewidth]{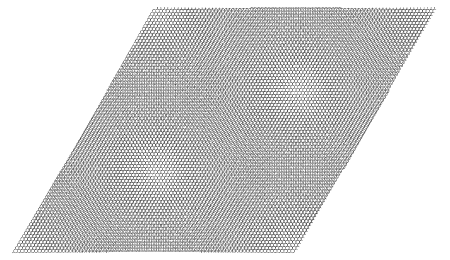}
}
	\hspace{-0.1\linewidth}
        \subfloat[]{
                \includegraphics[width=0.37\linewidth]{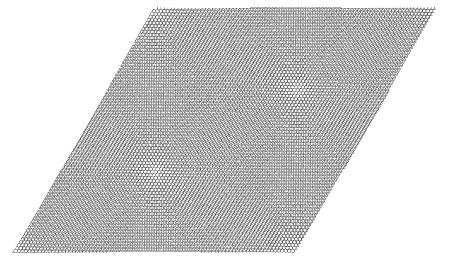}
        }
	\hspace{-0.1\linewidth}
        \subfloat[]{
                \includegraphics[width=0.37\linewidth]{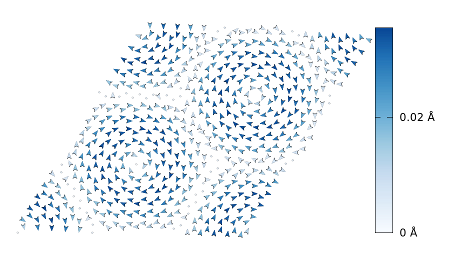}
        }
        \caption{The effects of relaxation are shown for a sample with (n,m) = (82,1), $\theta=1.2^\circ$ and $a_m$=115.3~\AA. (a) The sample prior to relaxation, (b) the sample after relaxation. (c) The displacements of the atoms as the result of relaxation for a sample (n,m) = (17,1), $\theta=5.7^\circ$ and $a_m$=24.5~\AA. The colour indicates size and the arrow the direction of the displacements.}
	\label{fig:relaxation}
\end{figure}

We first examine the case where the distance $z$ of the graphene layer to the rigid substrate is kept fixed and  atoms are allowed to move only in the $x$ and $y$ directions. The AA areas become smaller while the AB stacked areas expand, as shown by the difference between \fref{fig:relaxation}a and \fref{fig:relaxation}b. This effect is achieved by a small rotation of each moir\'e pattern around the AA center as illustrated in \fref{fig:relaxation}c where  we show  the displacements from the positions prior to relaxation. For clarity we show the displacement for a sample with $a_m=24.5$~\AA~but similar vortex-like structures occur also for larger moir\'e patterns.

\begin{figure}
        \centering
        \subfloat[$\theta=2.1^\circ$]{
	    \begin{minipage}{0.29\linewidth}
                \includegraphics[width=\linewidth]{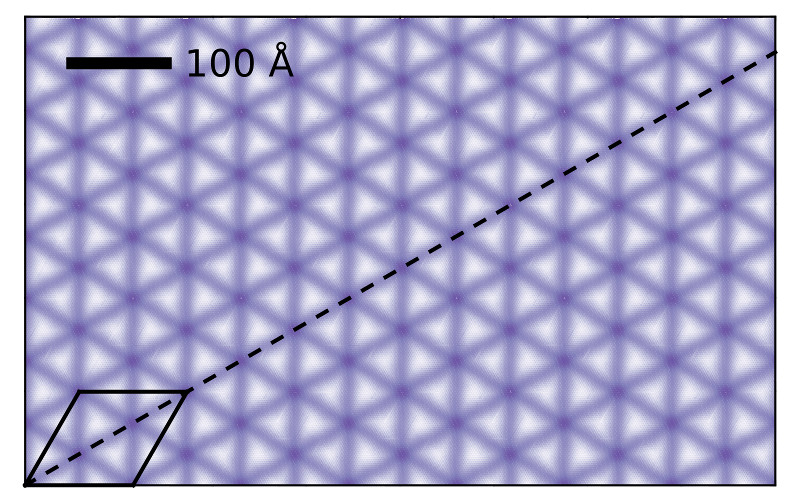}\\
		\includegraphics[width=\linewidth]{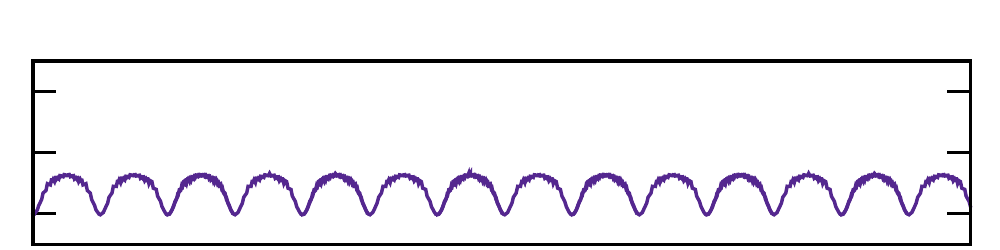}
	    \end{minipage}
	}
        \subfloat[$\theta=1.2^\circ$]{
	    \begin{minipage}{0.29\linewidth}
                \includegraphics[width=\linewidth]{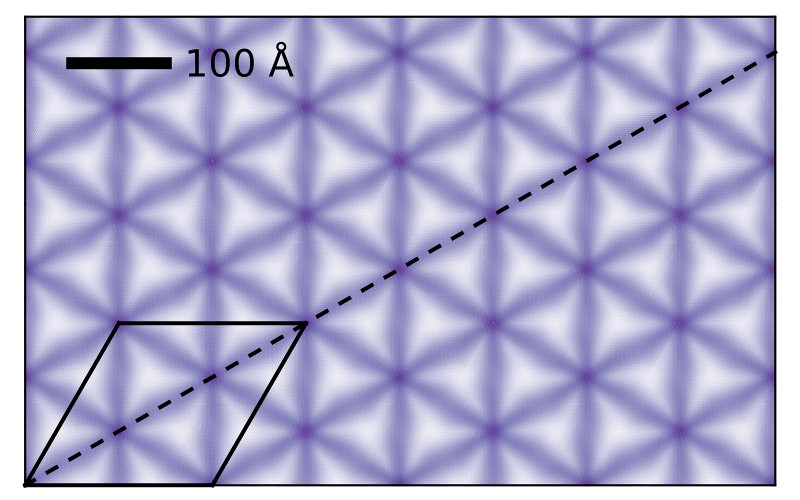}\\
		\includegraphics[width=\linewidth]{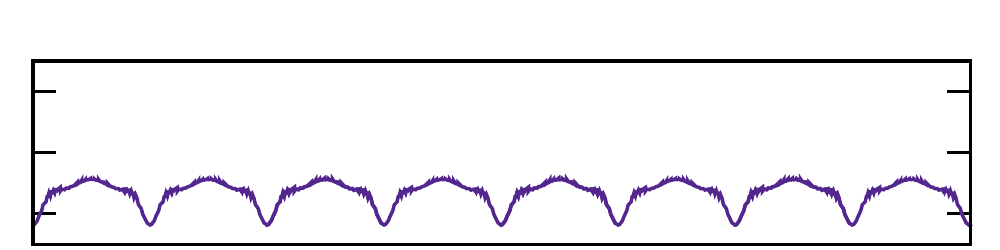}
	    \end{minipage}
	}
        \subfloat[$\theta=0.46^\circ$]{
	    \begin{minipage}{0.3806\linewidth}
                \includegraphics[width=\linewidth]{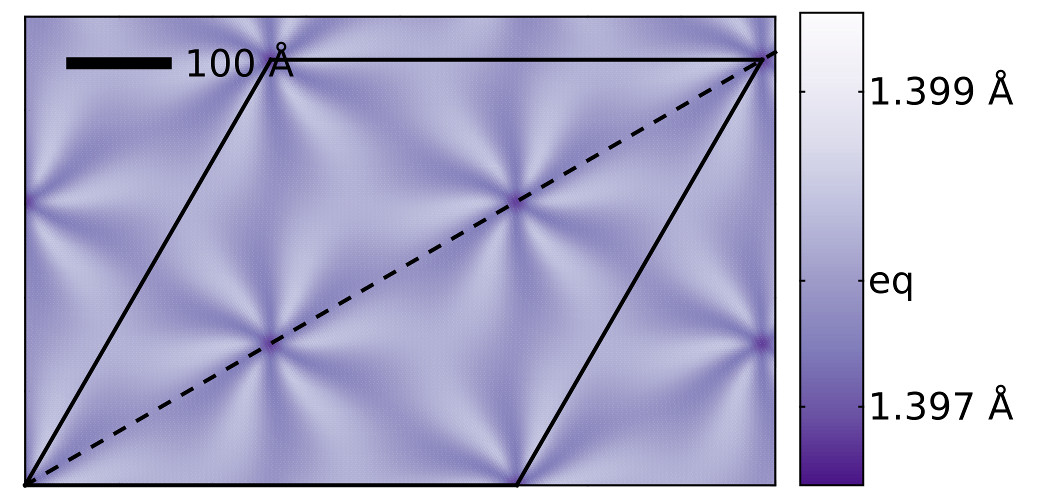}\\
		\includegraphics[width=\linewidth]{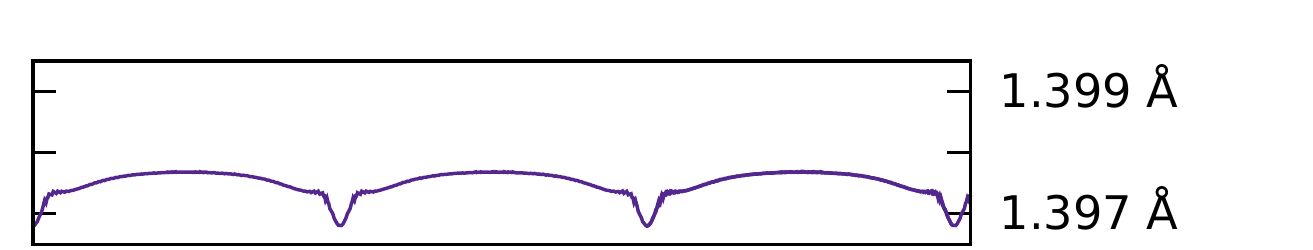}
	    \end{minipage}
	}
        \caption{Bondlenghts of relaxed configurations for samples with $z=3$~\AA. The supercell is shown in black. The bottom panels show the bond length along the dashed diagonal line. (a) $\theta=2.1^\circ$, (n,m)=(47,1), $a_m=$66.4~\AA. (b) $\theta=1.2^\circ$, (n,m)=(82,1), $a_m=$115.3~\AA. (c) $\theta=0.46^\circ$, (n,m)=(216,1), $a_m=$302.6~\AA.}
	\label{fig:dz3} 
\end{figure}

In \fref{fig:dz3} we show the distribution of bond lengths for decreasing angles. 
The rotational pattern in the displacements can also be observed in the bond lengths: the triangular pattern for $\theta=2.1^\circ$ and $\theta=1.2^\circ$ changes to a windmill shape for the larger moir\'e pattern with $\theta=0.46^\circ$. We see that by reducing the angle, the modulation of the bond length changes from an almost sinusoidal behaviour to rather localized contraction close to the AA areas. The `windmill' patterns appearing at small angles resemble those found in~\cite{snyman1981computed} as a function of the  misfit energy, which in our case would correspond to an increasing penalty for AA stacking. 

\begin{figure}
        \centering
        \subfloat[]{
                \includegraphics[width=0.48\linewidth]{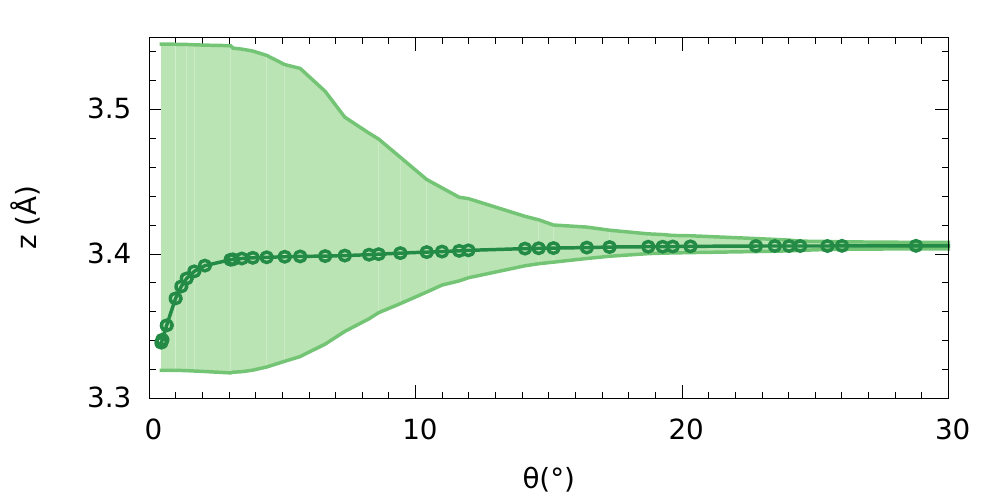}
        }
        \subfloat[]{
                \includegraphics[width=0.48\linewidth]{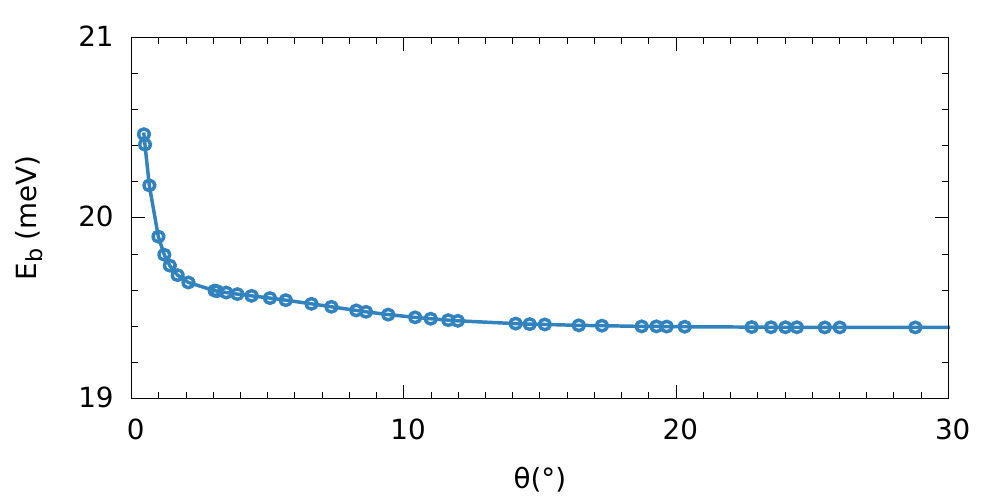}
        }
        \caption{(a) Average height of the layer relative to the other layer after relaxation for various samples. (b) Binding energy per atom after relaxation. For a rigid AA-stacked sample at the equilibrium distance $E_b$=16.96~meV and for AB-stacked $E_b$=21.32~meV.}
	\label{fig:zavandE}
\end{figure}

The distribution of bond lengths observed at fixed  $z=3$~\AA~is very similar to those found when atoms are free to move in $z$, although the differences are smaller than for $z=3$~\AA, due to the increased distance and hence smaller corrugation. When the atoms are allowed to relax in the out-of-plane ($z$) direction, this is the main direction of the atomic displacements. In \fref{fig:zavandE}a the average distance to the substrate $z$ is shown, together with the minimum and maximum values. For $\theta>20^\circ$, the graphene lies flat on the substrate. When $\theta$ decreases, the minimum and maximum values of $z$ are clearly different from the average. The spreading of $z$ agrees with the DFT calculations of ~\cite{uchida2014atomic} but the average value is not shown there. For $\theta<3^\circ$, the minimum and maximum value of z stay constant at the equilibrium values of the AA and AB stacking, but the average distance to the substrate decreases, implying that the AB stacked areas grow. The binding energy $E_b=(2 E_{graphene}-E)/N$, shown in \fref{fig:zavandE}b, stays constant for a large range of values as also found in~\cite{uchida2014atomic}. Below $\theta=2^\circ$, a range not reported in~\cite{uchida2014atomic}, we find that the binding energy increases significantly.

\begin{figure}
        \centering
        \subfloat[$\theta=2.1^\circ$]{
	    \begin{minipage}{0.29\linewidth}
                \includegraphics[width=\linewidth]{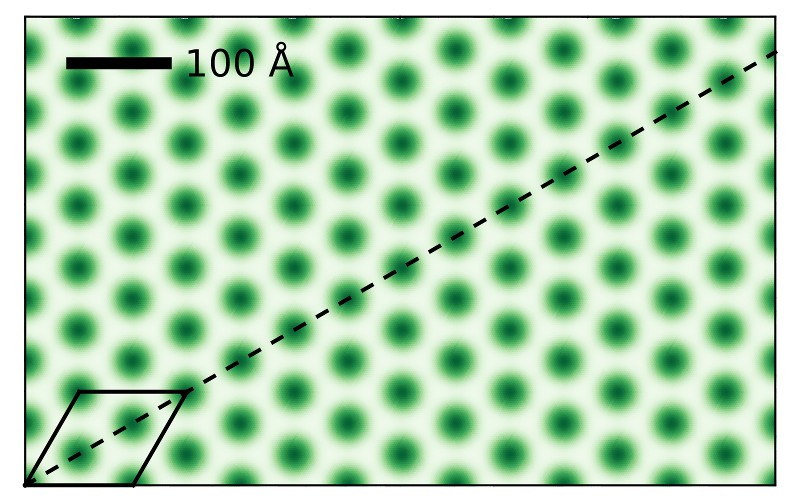}\\
		\includegraphics[width=\linewidth]{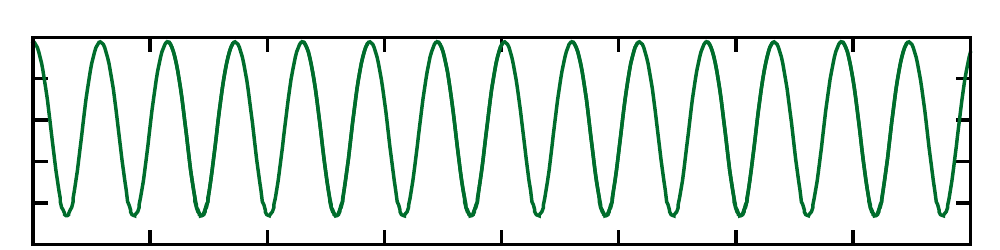}
	    \end{minipage}
	}
        \subfloat[$\theta=1.2^\circ$]{
	    \begin{minipage}{0.29\linewidth}
                \includegraphics[width=\linewidth]{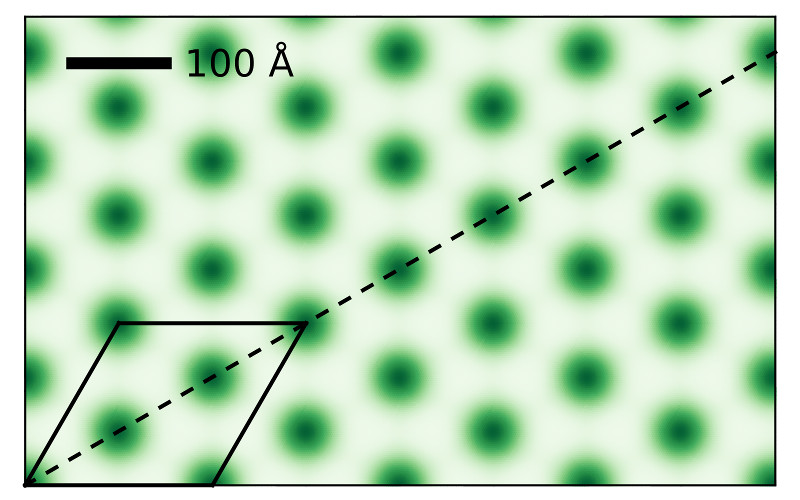}\\
		\includegraphics[width=\linewidth]{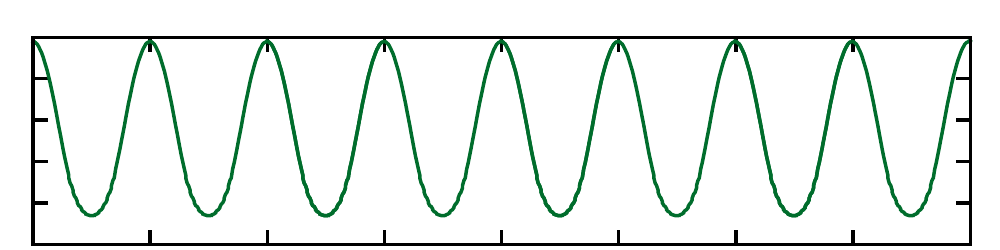}
	    \end{minipage}
	}
        \subfloat[$\theta=0.46^\circ$]{
	    \begin{minipage}{0.3806\linewidth}
                \includegraphics[width=\linewidth]{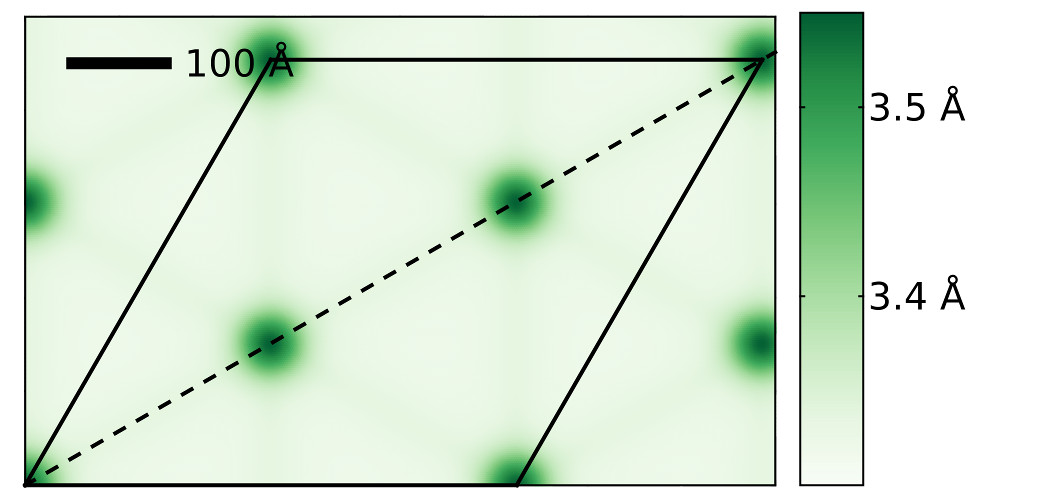}\\
		\includegraphics[width=\linewidth]{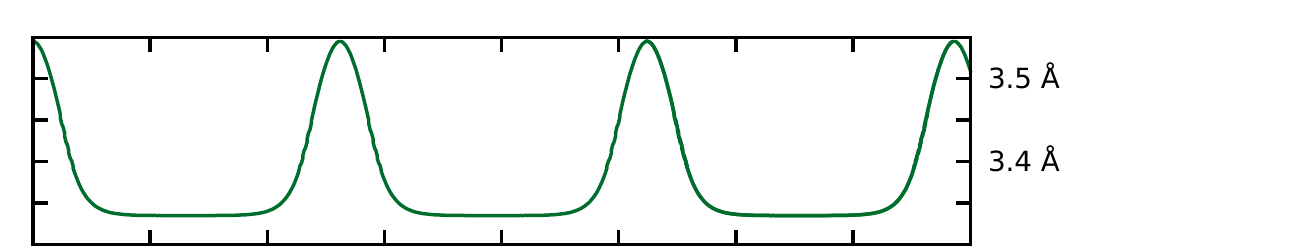}
	    \end{minipage}
	}
	\caption{Out-of-plane distance for samples where the graphene layer is relaxed in all dimensions. The bottom panels show the out-of-plane distance along the dashed diagonal line. (a) $\theta=2.1^\circ$, (n,m)=(47,1), $a_m=$66.4~\AA. (b) $\theta=1.2^\circ$, (n,m)=(82,1), $a_m=$115.3~\AA. (c) $\theta=0.46^\circ$, (n,m)=(216,1), $a_m=$302.6~\AA.}
	\label{fig:zfree}
\end{figure}

In \fref{fig:zfree} we show the distance from the substrate $z$ for three decreasing angles. In the panels showing the modulation along the cell diagonal, we see that the behaviour already found for the in-plane displacement is even more pronounced: for $\theta=2.1^\circ$ the modulation is roughly sinusoidal whereas for $\theta=0.46^\circ$ the `hot spots' are very pronounced and separated by flat domains. 
 
The main effects produced by structural deformations on the electronic properties of graphene are the modulation of the pseudo-electrostatic potential $V$ proportional to local uniform compression/dilatation and the appearance of a (pseudo)-vector potential $\vec{A}$ with components proportional to the shear deformations~\cite{vozmediano2010gauge, katsnelsonbook}:
\begin{equation}
\eqalign{V=2g\frac{\Delta d}{d}\cr
 A_x=\frac{\sqrt{3}}{2}\left(t_3-t_2\right)\approx \beta t \left(u_{xx}-u_{yy}\right)\cr
 A_y=\frac{1}{2}\left(t_2+t_3-2t_1\right)\approx -2\beta t u_{xy}}
\end{equation}
where $g\approx 3-4$ eV is the deformation potential for graphene, $\Delta d =\frac{1}{3}\left(d_1+d_2+d_3\right)-d$ with $d_1,d_2,d_3$ the first neighbour interatomic distances in the deformed lattice, $t_1,t_2,t_3$ are hopping parameters in the deformed lattice,
$t\approx 3$ eV is their unperturbed value, $\beta \approx 2-3$ is the electronic Gr\"uneisen parameter. 
The resulting pseudo-magnetic field $\vec{B}=\nabla \times \vec{A}$, which has an opposite sign for the two valleys K and K', can be estimated as~\cite{katsnelsonbook,morozov2006strong}
\begin{equation}
B \approx \frac{\hbar c}{e}\left(\frac{2}{3}\beta\right)\frac{\bar u}{d L}
\end{equation}
where ${\bar u}$ is a typical value of the shear deformation and $L$ is a typical size of the spatial variation of the deformation which we can estimate as $a_{m}$. 

To estimate the strength of the pseudo magnetic field, we calculate the magnitude of the shear deformations from the first neighbour interatomic distances in the deformed lattice as
\begin{equation}
{\bar u}=\frac{\sqrt{3\left(d_3-d_2\right)^2+\left( d_2+d_3-2d_1\right)^2}}{2d}.
\label{eq:ubar}
\end{equation}

\begin{figure}
        \centering
        \subfloat[$\theta=5.7^\circ$]{
                \includegraphics[width=0.48\linewidth]{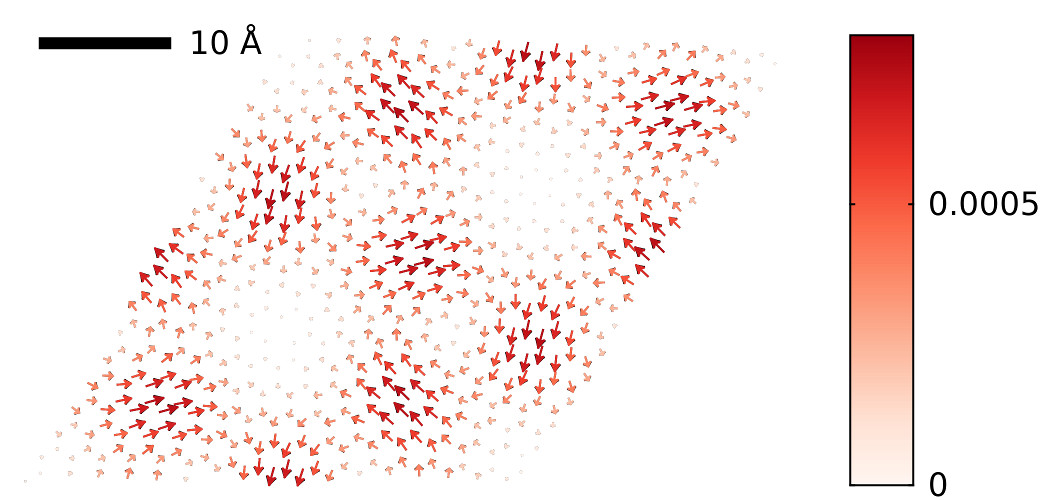}
        }
        \subfloat[$\theta=0.46^\circ$]{
                \includegraphics[width=0.48\linewidth]{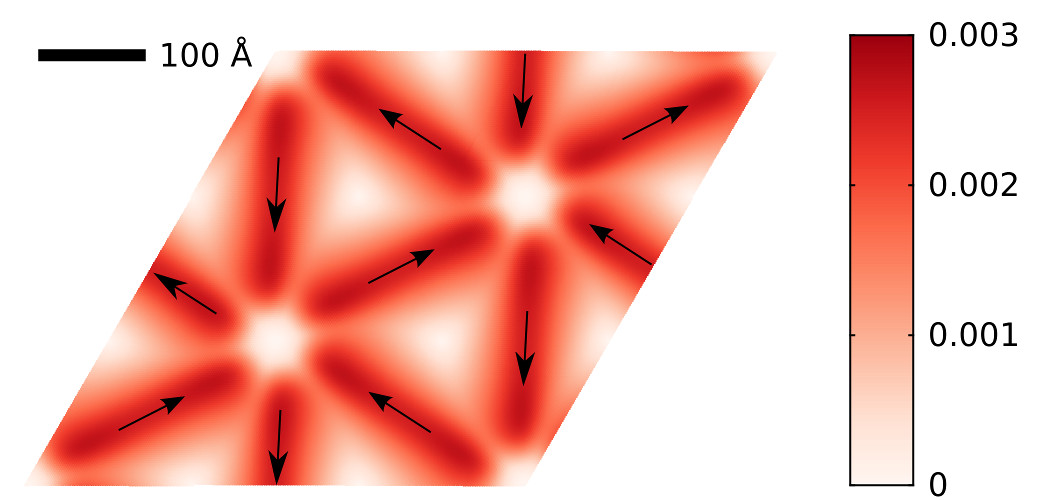}
        }
  \caption{${\bar u}$ as in equation \ref{eq:ubar} for (a)~$\theta=5.7^\circ$ and (b)~$\theta=0.46^\circ$.}
  \label{fig:vectorfield}
\end{figure}

Taking the parameters from \fref{fig:vectorfield}a,b we find in both cases $B$ of the order of 1~T, a value comparable to the one originating from  ripples for graphene in SiO$_2$~\cite{morozov2006strong}. The straightforward experimental evidence of the existence of this field would be the suppression of weak localization effects~\cite{morozov2006strong}. For the smallest angle the field is much less homogeneous. We have indicated the main directions of the vector potential in the areas of largest shear deformation by arrows. One can see that the pattern is not trivial and, for instance,  cannot be described as superposition of the field created by magnetic fluxes. The amplitude of the modulation of $V$ can be estimated to be a few meV. 

\section{Double layer graphene}
\begin{figure}
        \centering
        \subfloat[$\theta=2.1^\circ$]{
	    \begin{minipage}{0.29\linewidth}
                \includegraphics[width=\linewidth]{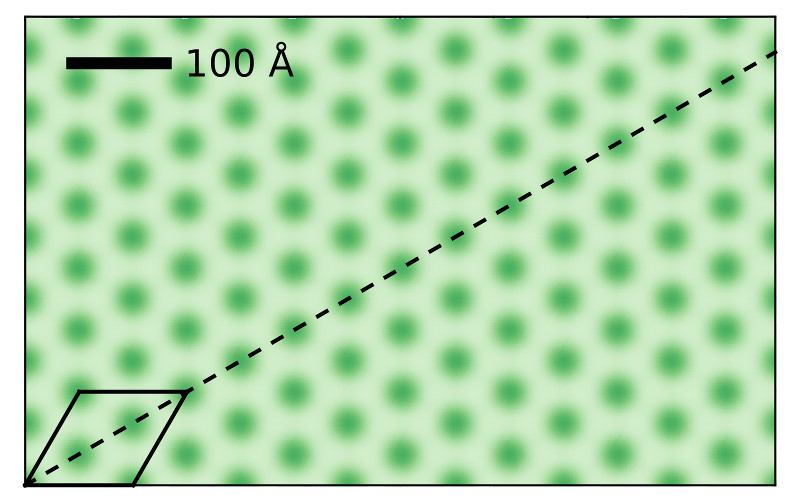}\\
		\includegraphics[width=\linewidth]{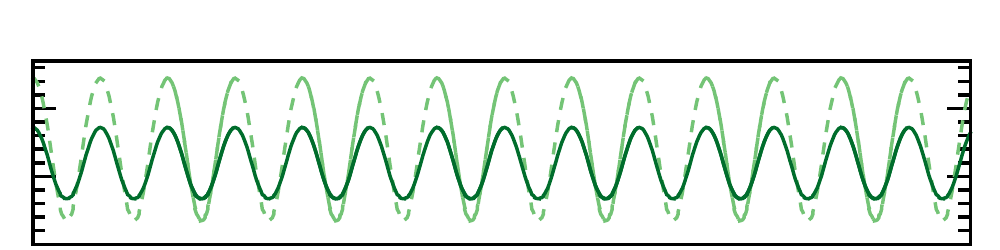}\\
		\includegraphics[width=\linewidth]{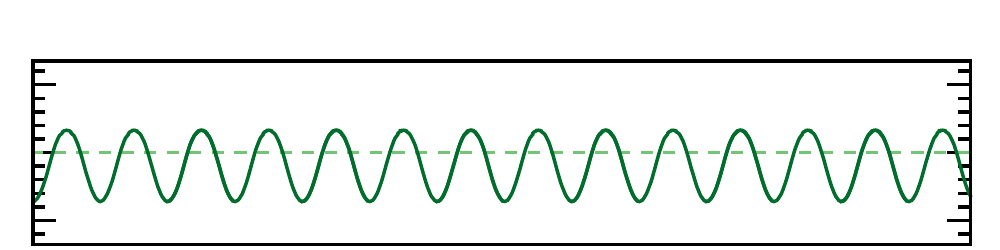}
	    \end{minipage}
	}
        \subfloat[$\theta=1.2^\circ$]{
	    \begin{minipage}{0.3806\linewidth}
                \includegraphics[width=\linewidth]{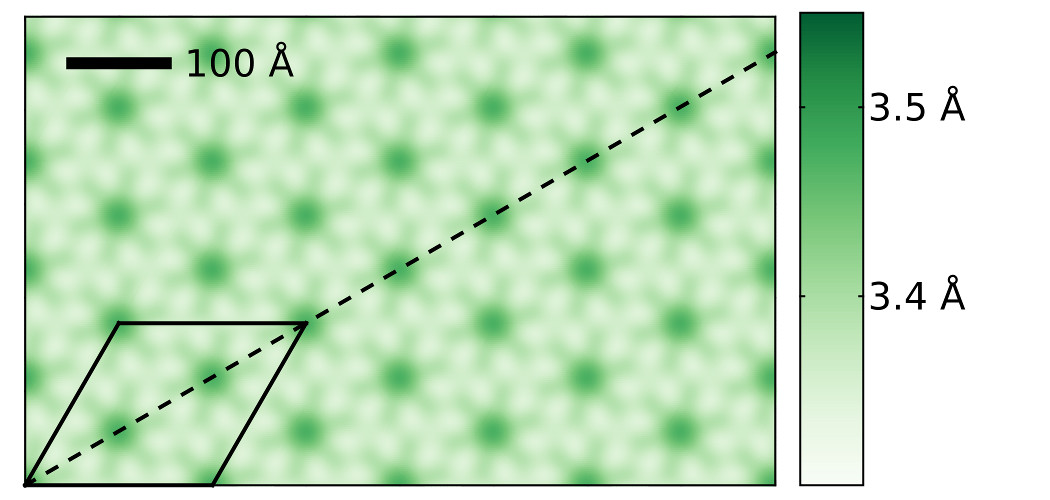}\\
		\includegraphics[width=\linewidth]{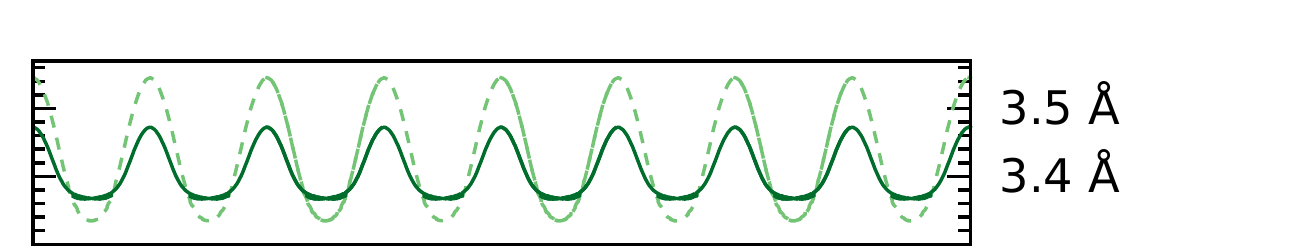}\\
		\includegraphics[width=\linewidth]{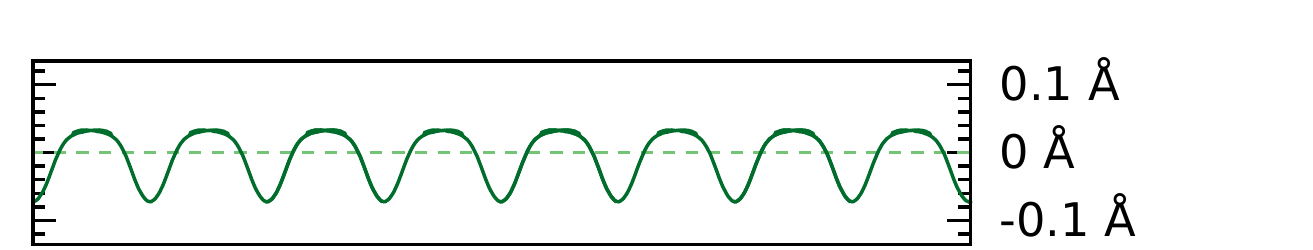}
	    \end{minipage}
	}
        \caption{Out-of-plane distance for double layer graphene. The bottom four panels show $z$ along the dashed line in the top figure. The dashed lines show the $z$ for graphene on graphite as in \fref{fig:zfree}.}
	\label{fig:doublelayer}
\end{figure}

In order to model double layer graphene, we no longer consider a rigid substrate, but two graphene layers which are both free to move in all directions. Whereas for graphene on graphite all corrugation occurs within one layer, in the case of a double layer the corrugation is shared between two layers, as shown in \fref{fig:doublelayer}. Furthermore we note that the deviation of the corrugation from the sinusoidal shape occurs for larger angles than in the case of graphene on graphite. Already at $\theta=1.2^\circ$ the behaviour is no longer sinusoidal and the relative deformation of the two layers seems to lead to a more complex pattern than for a single layer. 

\section{Conclusion}
We have shown that the lattice deformations in graphene on graphite and double layer graphene for small misorientation angle become very different from the sinusoidal modulation commonly used in theoretical models, leading to a structure of `hot spots' separated by roughly uniform domains. The relatively large out-of-plane deformation at the `hot spots' should be observable by scanning probe microscopy. Despite the modest effect of relaxation on the in-plane modulation, we estimate that it could give rise to a quite strong pseudomagnetic field, affecting the electronic structure of graphene. It would be interesting to pursue this topic further and study more precisely the effect of atomic relaxation of slightly misaligned graphene layers on the electronic properties.

\noindent
$ {\bf Acknowledgements} $. We acknowledge interesting discussions with Yury Gornostyrev and Kostya Novoselov. This work is part of the research program of the Foundation for Fundamental Research on Matter (FOM), which is part of the Netherlands Organisation for Scientific Research (NWO). The research leading to these results has received funding from the European Union Seventh Framework Programme under grant agreement n°604391 Graphene Flagship. This work is supported in part by COST Action MP1303.

\bibliographystyle{iopart-num}
\bibliography{refsGoGmoire_final}

\end{document}